\documentclass[pra,showpacs,superscriptaddress,amsmath,amssymb]{revtex4}
\usepackage{amsmath}
\usepackage{graphicx}
\usepackage[dvipdfm,pdfstartview=FitH]{hyperref}
\usepackage{bm}

\begin{document}

\title{Suppression of absorption by quantum interference in
intersubband transitions of tunnel-coupled double quantum wells}

\author{Wen-Xing Yang}
\email{wenxingyang2@126.com} \affiliation{Department of Physics,
Southeast University, Nanjing, Jiangsu 210096, China}
\affiliation{Institute of Photonics Technologies, National Tsing-Hua
University, Hsinchu 300, Taiwan}

\author{Jin Xu}
\affiliation{Department of Physics, Southeast University, Nanjing,
Jiangsu 210096, China}

\author{Ray-Kuang Lee}
\affiliation{Institute of Photonics Technologies, National Tsing-Hua
University, Hsinchu 300, Taiwan}
\date{\today}

\begin{abstract}

We propose and analyze an efficient scheme for suppressing the
absorption of a weak probe field based on intersubband transitions
in a four-level asymmetric coupled-quantum well (CQW) driven
coherently by a probe laser field and a control laser field. By
using the numerical simulation, we find that the magnitude of the
transient absorption of the probe field is smaller than the
three-level system based on the electromagnetically induced
transparency (EIT) at line center of the probe transition, moreover,
comparing with the scheme in three-level asymmetric double QW
system, the transparency hole of the present work is much broader.
By analyzing the steady-state process analytically and numerically,
our results show that the probe absorption can be completely
eliminated under the condition of Raman resonance (i.e. two-photon
detuning is zero). Besides, we can observe one transparency window
without requiring one- or two-photon detuning exactly vanish. This
investigation may provide the possible scheme for EIT in solids by
using CQW.

\end{abstract}


\maketitle

\section{Introduction}
The optical properties of cold atomic gases can be radically
modified by laser beams\cite{1}. A strong laser light essentially
influences atomic states, quantum coherence and interference between
the excitation pathways controlling the optical response become
possible, and the absorption of weak laser light vanishes. This
effect was termed electromagnetically induced transparency
(EIT)\cite{2}. The phenomenon of EIT has been deeply studied in
atomic physics\cite{1}-\cite{10}, starting from its observation in
sodium vapors\cite{11}, where the effect of EIT in atomic system has
disclosed new possibilities for nonlinear optics and quantum
information processing. For example, few works have demonstrated
that EIT can be used to suppress both single-photon and two-photon
absorptions in three-level atomic medium\cite{6}. The schemes for
suppressing absorption of the short-wavelength light based on EIT
have also been proposed\cite{9}. In particular, few authors analyzed
and discussed suppressed both two- and three-photon absorptions in
four-wave mixing (FWM) and hyper-Raman scattering (HRS) in resonant
coherent atomic medium via EIT\cite{7,8,9}.

As we know, it is easy to implement EIT in optically dense atomic
medium in gases phase, but it is more difficult to observe EIT in
solid-state medium due to the short coherence times in solid-state
system\cite{1}. Nevertheless, some investigations have
demonstrated EIT effect and ultraslow optical pulse propagation in
semiconductor quantum wells (QW) structure\cite{12}-\cite{16}. For
example, Sadeghi et al.\cite{13}, in an asymmetric QW, have shown
that EIT may be obtained with appropriate driving fields, provided
that the coherence between the intersubbands considered is
preserved for a sufficiently long time. So that quantum coherence
and interference in QW structures have also attracted great
interesting due to the potentially important applications in
optoelectronics and solid-state quantum information science. In
fact, except for the investigations of EIT, Autler-Townes
splitting\cite{17}, gain without inversion\cite{18}, modified Rabi
oscillations and controlled population transfer\cite{19,20},
largely enhanced second harmonic generation\cite{21}, controlled
optical bistability\cite{22}, ultrafast all-optical
switching\cite{23}, phase-controlled behavior of absorption and
dispersion\cite{24}, coherent population trapping\cite{25},
enhanced refraction index without absorption\cite{26},
FWM\cite{27}, and several other phenomena\cite{28}-\cite{30}, have
been theoretically studied and experimentally observed in
intersubband transitions of semiconductor QW. In addition, we take
notice of the calculation of the model for an asymmetric
semiconductor coupled double quantum well\cite{31}-\cite{34}, in
which the realization of a photon switch and the optical
bistability have been studied based on quantum interference
between intersubband transitions in this system\cite{32}. The aim
of our current study is to analyze and discuss the probe
absorption properties for the two cases of transient- and
steady-state processes in such system. Different from the
intersubband transition in infrared region, we consider such
transition in the ultraviolet or visible region, in which it has
very large dipole moment length, so the optoelectonics devices may
be realized with relatively smaller pulse intensity. Besides,
Devices based on semiconductor QW structures have many inherent
advantages, such as large electric dipole moments due to the small
effective electron mass, high nonlinear optical coefficients, and
a great flexibility in device design by choosing the materials and
structure dimensions.

\section{The physical model and equations of motion}
\begin{figure}
\includegraphics[width=9cm]{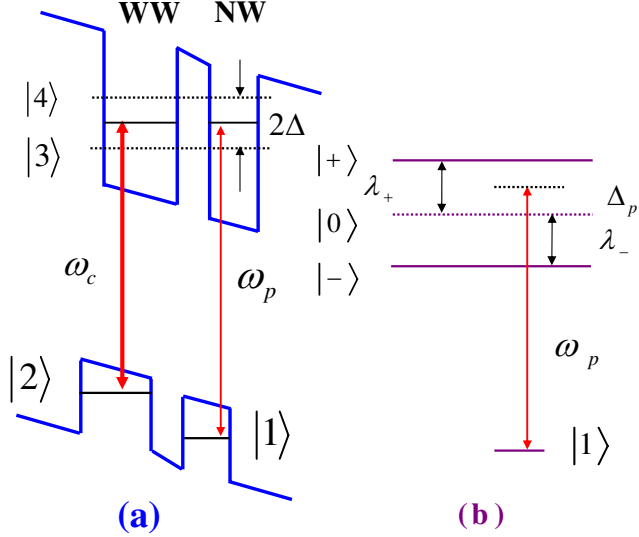}
\caption{\label{fig1} Conduction subband energy level diagram for an
asymmetric coupled quantum wells consisting of a wide well (WW) and
a narrow well (NW), as shown in Ref.\cite{33}. (a) Levels $\left| 3
\right\rangle $ and $\left| 4 \right\rangle $ are the bonding and
antibonding electron state of coupled system and Levels $\left| 1
\right\rangle $ and $\left| 2 \right\rangle $ are the localized hole
states. $2\Delta$, $\omega_c$ and $\omega_p$ are the energy
splitting between tunnel-coupled electronic levels, frequency of the
strong coupled field and the weak probe field, respectively. (b)
Corresponding to the dressed-state picture for the case
$\Delta_c=0$: subband $\left| 2 \right\rangle $ couple with subbands
$\left| 3 \right\rangle $ and $\left| 4 \right\rangle $
simultaneously giving rise to the three dressed states $\left| +
\right\rangle $, $\left| 0 \right\rangle $ and $\left| -
\right\rangle $.}
\end{figure}

Let us consider the asymmetric semiconductor coupled double quantum
well structure consisting of 10 pairs of a 51-monolayer (145{\AA})
thick wide well and a 35-monolayer (100{\AA})thick narrow well,
separated by a Al$_{0.2}$Ga$_{0.8}$As buffer layer\cite{33,34}, as
shown in Fig. \ref{fig1}. In this quantum well structure, the first
($n=1)$ electron level in the wide well and the narrow well can be
energetically aligned with each other by applying a static electric
field, while the corresponding $n=1$ hole levels are never aligned
for this polarity of the field. For transitions from the wide well
and the narrow well, the Coulomb interaction between electron and
the hole downshifts the value of the electric field where the
resonance condition is fulfilled to that corresponding to the
built-in field. Level $\left| 3 \right\rangle $ and level $\left| 4
\right\rangle $, respectively corresponding the bonding state and
the anti-bonding state, are tunnel-coupled new states of the first
($n=1)$ electron level in the wide well and the narrow well. The
energy difference $2\Delta$ of the bonding state and anti-bonding
state is determined by the level splitting in the absence of
tunneling and the tunneling matrix element, and can be controlled by
an electric field applied perpendicularly to coupled-quantum wells.
Level $\left| 1 \right\rangle $ in the narrow well and level $\left|
2 \right\rangle $ in the wide well are localized hole states.

We assume the transitions $\left| 2 \right\rangle \leftrightarrow
\left| 4 \right\rangle $ and $\left| 2 \right\rangle \leftrightarrow
\left| 3 \right\rangle $ are simultaneously driven by a strong
coupling field with the respective one-half Rabi frequencies $\Omega
_c = \mu _{42} E_c / 2\hbar $ and $(\Omega _c \mu _{32} ) / \mu
_{42} $. At the same time, a weak probe field is applied to the
transitions $\left| 1 \right\rangle \leftrightarrow \left| 4
\right\rangle $ and $\left| 1 \right\rangle \leftrightarrow \left| 3
\right\rangle $ simultaneously with the respective Rabi frequencies
$\Omega _p = \mu _{41} E_p / 2\hbar $ and $(\Omega _p \mu _{31} ) /
\mu _{41} $. $E_c $ and $E_p $ are the amplitude of the
strong-coupling field and the weak probe field, respectively. $\mu
_{ij} = \mu _{ij} \cdot \widehat{e}_{L}$ is the dipole moment for
the transition between levels $\left| i \right\rangle $ and $\left|
j \right\rangle $ with $\widehat{e}_{L} $ being the unit
polarization vector of the corresponding laser field. In
schr\"odinger picture and in rotating-wave approximation (RWA), the
semiclassical Hamiltonian describing the system under study can
present as

\begin{equation}
\label{eq0} H = \sum\limits_{j = 1}^4 {E_j \left| j \right\rangle
\left\langle j \right|} - \hbar (k\Omega _c e^{ - i\omega _c
t}\left| 3 \right\rangle \left\langle 2 \right| + \Omega _c e^{ -
i\omega _c t}\left| 4 \right\rangle \left\langle 2 \right| + q\Omega
_p e^{ - i\omega _p t}\left| 3 \right\rangle \left\langle 1 \right|
+ \Omega _p e^{ - i\omega _p t}\left| 4 \right\rangle \left\langle 1
\right| + H.c.).
\end{equation}

\noindent Here $E_j = \hbar \omega _j $ is the energy of the subband
level $\left| j \right\rangle $ ($j = 1,2,3,4)$, and the parameters
$k=\mu_{32}/\mu_{42}$, $q=\mu_{31}/\mu_{41}$ represent the ratios
between the dipole moments of the subband transitions. Working in
the interaction picture, utilizing RWA and the electric-dipole
approximation (EDA), we derive the interaction Hamiltonian for the
system as (assuming $\hbar=1$)

\begin{equation}
\label{eq1} H_I = (\Delta _c - \Delta _p )\left| 2 \right\rangle
\left\langle 2 \right| - (\Delta _p + \Delta)\left| 3 \right\rangle
\left\langle 3 \right| + (\Delta - \Delta _p )\left| 4 \right\rangle
\left\langle 4 \right| + (\Omega _c \left| 4 \right\rangle
\left\langle 2 \right| + k\Omega _c \left| 3 \right\rangle
\left\langle 2 \right| + \Omega _p \left| 4 \right\rangle
\left\langle 1 \right| + q\Omega _p \left| 3 \right\rangle
\left\langle 1 \right| + H.c.),
\end{equation}

\noindent with $H_0=(\omega_p-\omega_c)\left| 2 \right\rangle
\left\langle 2 \right| + \omega_p\left| 3 \right\rangle \left\langle
3 \right| +\omega_p\left| 4 \right\rangle \left\langle 4\right|$,
$\Delta _c = \omega _c + \omega _2 - \omega _0 $, $\Delta _p =
\omega _p - \omega _0 $, $\omega _0 = \frac{1}{2}(\omega _4 + \omega
_3 )$ and $\Delta = (E_4 - E_3)/2 $. $\Delta _p $ is the detuning
between the frequency of the probe field and the average transition
frequency $\omega _0 $, $\Delta _c $ is the detuning between the
frequency of the strong-coupling field and the average frequency. It
should be emphasized that the electron sheet density of the quantum
well structure is such that electron-electron effects have very
small influence in our results. Therefore, the effects of
electron-electron interactions are not considered in this present
work. Besides, we have regarded the $E_1 $ as the energy origin. For
simplicity, we assume that the Rabi frequency of external field
$\Omega_c$ and $\Omega_p$ are real. Based on the equation of motion
for density operator in the interaction picture
($\partial\rho/\partial t = -i[H_I, \rho]+\Lambda\rho)$, we can
easily derive the density matrix equations of motion as follows

\begin{equation}
\label{eq2} \frac{\partial \rho _{44} }{\partial t} = i\Omega _c
(\rho _{24} - \rho _{42} ) + i\Omega _p (\rho _{14} - \rho _{41} ) -
(2\gamma _{41} + 2\gamma _{42} )\rho
_{44}-\frac{\kappa}{2}(\rho_{34}+\rho_{43}),
\end{equation}

\begin{equation}
\label{eq3} \frac{\partial \rho _{33} }{\partial t} = ik\Omega _c
(\rho _{23} - \rho _{32} ) + iq\Omega _p (\rho _{13} - \rho _{31} )
- (2\gamma _{31} + 2\gamma _{32} )\rho
_{33}-\frac{\kappa}{2}(\rho_{34}+\rho_{43}),
\end{equation}

\begin{equation}
\label{eq4} \frac{\partial \rho _{22} }{\partial t} = ik\Omega _c
(\rho _{32} - \rho _{23} ) + i\Omega _p (\rho _{42} - \rho _{24} ) -
2\gamma _{21} \rho _{22} + 2\gamma _{32} \rho _{33} + 2\gamma _{42}
\rho _{44},
\end{equation}

\begin{equation}
\label{eq5} \frac{\partial \rho _{34} }{\partial t} = i\omega _s
\rho _{34} + ik\Omega _c \rho _{24} + iq\Omega _p \rho _{14} -
i\Omega _c \rho _{32} - i\Omega _p \rho _{31} - \Gamma _{34}\rho
_{34}-\frac{\kappa}{2}(\rho_{44}+\rho_{33}),
\end{equation}

\begin{equation}
\label{eq6} \frac{\partial \rho _{24} }{\partial t} = i(\Delta -
\Delta _c )\rho _{24} + i\Omega _c (\rho _{44} - \rho _{22} ) +
ik\Omega _c \rho _{34} - i\Omega _p \rho _{21} - \Gamma _{24}\rho
_{24}-\frac{\kappa}{2}\rho_{23},
\end{equation}

\begin{equation}
\label{eq7} \frac{\partial \rho _{14} }{\partial t} = i(\Delta -
\Delta _p )\rho _{14} + i\Omega _p (\rho _{44} - \rho _{11} ) +
iq\Omega _p \rho _{34} - i\Omega _c \rho _{12} - \Gamma _{14}\rho
_{14}--\frac{\kappa}{2}\rho_{13},
\end{equation}

\begin{equation}
\label{eq8} \frac{\partial \rho _{23} }{\partial t} = -i(\Delta +
\Delta _c )\rho _{23} - iq\Omega _p \rho _{21} + i\Omega _c \rho
_{43}
 + ik\Omega _c
(\rho _{33} - \rho _{22} ) - \Gamma _{23}\rho
_{23}-\frac{\kappa}{2}\rho_{13},
\end{equation}

\begin{equation}
\label{eq9} \frac{\partial \rho _{13} }{\partial t} = - i(\Delta +
\Delta _p )\rho _{13} + i\Omega _p \rho _{43} - ik\Omega _c \rho
_{12} + iq\Omega _p (\rho _{33} - \rho _{11} ) - \Gamma _{13}\rho
_{13}-\frac{\kappa}{2}\rho_{24},
\end{equation}

\begin{equation}
\label{eq10} \frac{\partial \rho _{12} }{\partial t} = i(\Delta _c -
\Delta _p )\rho _{12} - ik\Omega _c \rho _{13} + iq\Omega _p \rho
_{32}
 - i\Omega _c \rho _{14} + i\Omega _p \rho
_{42} - \Gamma _{12}\rho _{12}-\frac{\kappa}{2}\rho_{14},
\end{equation}

\noindent together with $\rho _{ij} = \rho _{ji}^\ast $ and the
carrier conservation condition $\rho _{11} + \rho _{22} + \rho _{33}
+ \rho _{44} = 1$. Here the population decay rates $\gamma _{il} $
(life broadening) and the dephasing decay rates $\gamma _{ij}^{dph}
$ (dephasing line width corresponding to the respective transitions)
are added phenomenologically in the above equations with the total
decay rates (line width) being $\Gamma _{12} = \gamma _{12}^{dph} +
\frac{\gamma_{2l}}{2}$, $\Gamma _{13} = \gamma _{13}^{dph} +
\frac{\gamma_{3l}}{2}$, $\Gamma _{14} = \gamma _{14}^{dph}+
\frac{\gamma_{4l}}{2} $, $\Gamma _{23} = \gamma _{23}^{dph}+
\frac{\gamma_{2l}+\gamma_{3l}}{2}$, $\Gamma _{24} = \gamma
_{24}^{dph} + \frac{\gamma_{2l}+\gamma_{4l}}{2}$, and $\Gamma _{34}
= \gamma _{34}^{dph} + \frac{\gamma_{3l}+\gamma_{4l}}{2}$, where
$\gamma _{il} $ is determined by longitudinal optical (LO) phonon
emission events at low temperature, the $\gamma _{ij}^{dph} $ is
determined by electron-electron, interface roughness, and phonon
scattering processes\cite{16}. The population decay rates can be
calculated, and as we know, the initially nonthermal carrier
distribution is quickly broadened due to inelastic carrier-carrier
scattering, with the broadening rate increasing as carrier density
is increased. For the temperatures up to 10 K, the carrier density
smaller than $10^{11}$ cm$^{-2}$, the dephasing decay rates
$\gamma^{dph}_{ij}$ can be estimated according to Ref.\cite{23}.
Besides, the dephasing-type broadening
$\kappa=\sqrt{\gamma_{3l}\gamma_{4l}}$ denotes the cross coupling of
states  $\left| 4 \right\rangle$ and $ \left| 3 \right\rangle$ via
the LO phonon decay. A more complete theoretical treatment taking
into account these process for the dephasing rates is though
interesting but beyond the scope of this paper.

In the following analysis, we choose the parameters are set to be
$\mu _{32} = \mu _{42} = \mu _c $ and $\mu _{31} = \mu _{41} = \mu
_p $ (i.e. $k=1$ and $q=1$); and we take $\Gamma_{12}=0$ for the
lifetime of levels $\left| 2 \right\rangle$ and $\left| 1
\right\rangle$ according to the Ref.\cite{35}. By a straightforward
semiclassical analysis, the above matrix elements can be used to
calculate the total linear complex susceptibility $\chi $ of the
probe transitions $\left| 1 \right\rangle\rightarrow\left| 3
\right\rangle$ and $\left| 1 \right\rangle\rightarrow\left| 4
\right\rangle$, i.e.

\begin{equation}
\label{eq11} \chi = \frac{N\left| {\mu _p } \right|^2(\rho _{13} +
\rho _{14} )}{2\hbar \varepsilon _0 \Omega _p },
\end{equation}

\noindent where $N$ is the electron number density in the medium and
$\varepsilon _0 $ is the permittivity of free space. Based on the
Eq. (\ref{eq11}), one can find that the gain-absorption coefficient
for the probe field coupled to the transitions $\left| 1
\right\rangle \leftrightarrow \left| 3 \right\rangle $ ($\left| 4
\right\rangle )$ is proportional to the term Im$(\rho _{13} + \rho
_{14} )$ in the limit of a weak probe field. Im$(\rho _{13} + \rho
_{14} )> 0$ presents the absorption of the probe field, Im$(\rho
_{13} + \rho _{14} ) < 0$ presents the amplified of the probe field.
In the following, we investigate the absorption properties of such a
weak probe field for the transient- and steady-state processes. We
begin by solving numerically the time-dependent differential
equations (\ref{eq2})-(\ref{eq10}) by using Runge-Kutta algorithm
with the initial conditions $\rho _{11} (0) = 1$, $\rho _{22} (0) =
\rho _{33} (0) = \rho _{44} (0) = 0$ and $\rho _{ij} (0) = 0$ for $i
\ne j$ ($i,j = 1,2,3,4)$. Note that, the parameters $\Omega _c $,
$\Omega _p $, $\Delta _c $, $\Delta _p $, $\Delta$, $\gamma _{ij} $,
$\gamma _{ij}^{dph} $ are defined as units of $\gamma $ in this
present paper.

\section{The probe absorption under the transient-state process}

\begin{figure}\label{fig2}
\includegraphics[width=7cm]{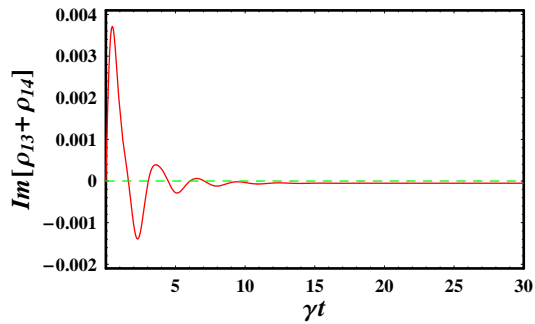}
\caption{\label{fig2} Transient absorption and amplification
response for a probe laser field with parameters being
$\Omega_p=0.01\gamma$, $\Delta=\gamma$, $\Delta_c=\Delta_p=0$,
$\gamma_{3l}=\gamma_{4l}=0.4\gamma$,
$\gamma_{14}^{dph}=\gamma_{24}^{dph}=\gamma_{13}^{dph}=\gamma_{23}^{dph}=0.1\gamma$.}
\end{figure}

Firstly, we examined the probe absorption under the transient-state
process by using different sets of parameters.

We show in Fig.\ref{fig2} the time evolution of the absorption
coefficient Im$(\rho _{13} + \rho _{14} )$ under the Raman resonant
condition $\Delta_c=\Delta_p$ with the parameters-value
$\Omega_p=0.01\gamma$, $\Omega_c=4\gamma$, $\Delta=\gamma$,
$\gamma_{3l}=\gamma_{4l}=0.4\gamma$ and
$\gamma_{14}^{dph}=\gamma_{24}^{dph}=\gamma_{13}^{dph}=\gamma_{23}^{dph}=0.1\gamma$.
The probe field shows the oscillatory behavior versus the time
before reaching the steady state, which just like in high density
atomic system. The probe absorption increases rapidly to a maximum
peak value, then it decreases gradually to a peak value and again
increases with increase of the evolution time, up to a negative
large steady-state value below the zero-absorption line. Comparing
with the three-level EIT system, the transient behavior of the probe
field can be tuned by adjusting the coherent tuneling (coupling)
between the two upper levels $\left| 3 \right\rangle $ and $\left| 4
\right\rangle$ across a thin barrier, we can expect that the
coupling strength, i.e., the energy splitting $2\Delta$, must play
an important role in controlling the transient behavior of the probe
field in such an asymmetric double QW system.

\begin{figure}\label{fig3}
\includegraphics[width=7cm]{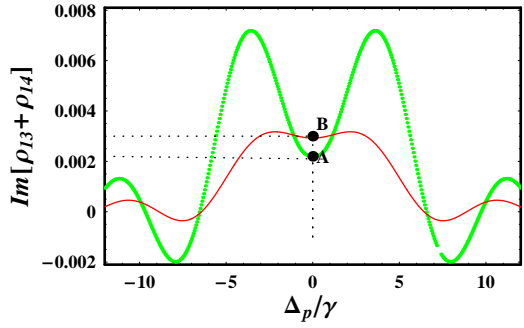}
\caption{\label{fig4} The probe absorption as a function of
dimensionless probe detuning $\Delta_p/\gamma$ for the transition
under the condition of transient-state process. Solid red curve:
$\Delta=0$; dot green curve: $\Delta=\gamma$. Other fitting
parameters used are $\Omega_{c}=2\gamma$, $\Omega_p=0.01\gamma$,
$\Delta_c=0$, $\gamma_{3l}=\gamma_{4l}=0.4\gamma$,
$\gamma_{14}^{dph}=\gamma_{24}^{dph}=\gamma_{13}^{dph}=\gamma_{23}^{dph}=0.1\gamma$.}
\end{figure}

\begin{figure}\label{fig4}
\includegraphics[width=7cm]{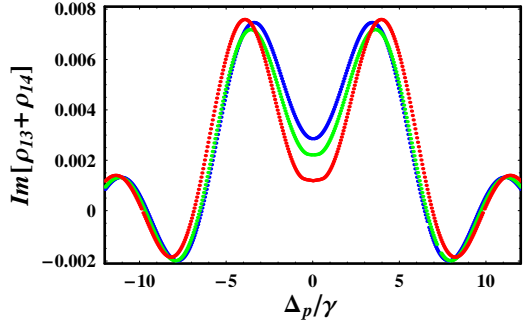}
\caption{\label{fig4} Imaginary part of the susceptibility of the
probe transition as a function of dimensionless detuning
$\Delta_{p}/\gamma$ under the condition of transient-state process
with $\Omega_{c}=1.5\gamma$ (blue line), $\Omega_{c}=2\gamma$ (green
line) and $\Omega_{c}=1\gamma$ (red line). Other parameters are
$\Omega_p=0.01\gamma$, $\Delta=\gamma$, $\Delta_c=0$,
$\gamma_{3l}=\gamma_{4l}=0.4\gamma$,
$\gamma_{14}^{dph}=\gamma_{24}^{dph}=\gamma_{13}^{dph}=\gamma_{23}^{dph}=0.1\gamma$.}
\end{figure}

We show in Fig.3 the magnitudes of the transient absorption of the
probe field versus the probe detuning $\Delta_p/\gamma$. At the line
center of the probe transition $\Delta_p=0$, the magnitude of the
transient absorption of the probe field at point A corresponding to
the splitting $2\Delta$ is much smaller than the magnitude of the
probe field at point B for the three-level EIT corresponding to
$\Delta=0$. With appropriate values of parameters we obtain the
absorption spectra shown in Fig.\ref{fig4} for different intensities
of the strong-coupling field. With the increasing the intensity of
the strong-coupling field, the absorption deep of the probe field
moves small accordingly, as expected for coherent EIT. At the same
time, with a larger intensity of the control field the transparency
width increases, which corresponds to the splitting between the two
upper levels $\left| 3 \right\rangle$ and $\left| 4 \right\rangle$.
From the above analysis, we can conclude that the probe absorption
at the line center ($\Delta _p = 0)$ can be considerably suppressed
due to the fact of the splitting $2\Delta$.

\begin{figure}\label{fig5}
\includegraphics[width=8cm]{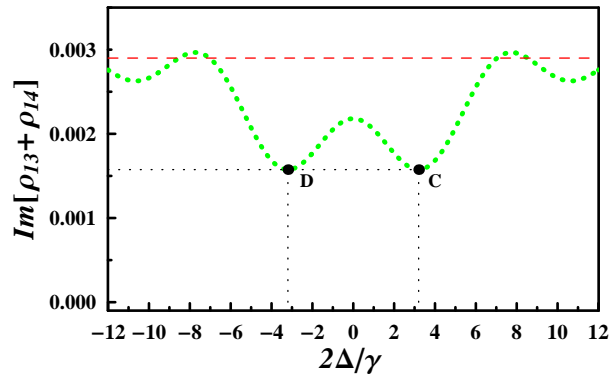}
\caption{\label{fig5} The probe absorption as a function of
dimensionless splitting  $\Delta/\gamma$ between $\left| 3
\right\rangle$ and $\left| 4 \right\rangle$ under the condition of
transient-state process (green curve). The dashed red line stands
for the magnitude of the probe absorption for the three-level EIT at
$\Delta_p=\Delta_c=0$. Other fitting parameters used are
$\Delta_p=0$, $\Delta_c=0$,$\Omega_p=0.01\gamma$, $\Omega_c=2\gamma$
$\gamma_{3l}=\gamma_{4l}=0.4\gamma$,
$\gamma_{14}^{dph}=\gamma_{24}^{dph}=\gamma_{13}^{dph}=\gamma_{23}^{dph}=0.1\gamma$.}
\end{figure}

\begin{figure}\label{fig6}
\includegraphics[width=8cm]{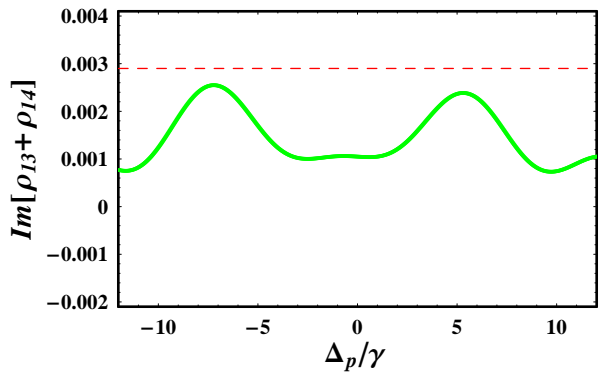}
\caption{\label{fig6} The probe transition as a function of
dimensionless detuning $\Delta_{p}/\gamma$ under the condition of
transient-state process for the Ramman resonance $\Delta_c=\Delta_p$
(green curve). The dashed line stands for the magnitude of the probe
absorption for three-level EIT at $\Delta_c=\Delta_p=0$. Other
fitting parameters being $\Omega_p=0.01\gamma$, $\Omega_c=2\gamma$,
$\Delta=\gamma$, $\gamma_{3l}=\gamma_{4l}=0.4\gamma$,
$\gamma_{14}^{dph}=\gamma_{24}^{dph}=\gamma_{13}^{dph}=\gamma_{23}^{dph}=0.1\gamma$.}
\end{figure}

The calculated probe absorption coefficient of such a weak probe
laser versus the splitting ($2\Delta$) between $\left| 3
\right\rangle$ and $\left| 4 \right\rangle$ for the probe and
coupling laser detunings $\Delta_p=\Delta_c=0$ is shown in
Fig.\ref{fig5}, in which the dashed red line stands for the
magnitude of the probe absorption for the standard three-level EIT
and the positions C and D correspond to the minimum of the probe
absorption. We observe that the magnitude of the transient
absorption of the probe field with the inclusion of the splitting
$2\Delta$ is almost lower than the magnitude of the probe absorption
with the three level EIT at $\Delta_p=\Delta_c=0$. As shown in
Fig.\ref{fig6}, we plot absorption profiles versus the probe
detuning $\Delta _p $ in the case of two-photon Raman resonance
$\Delta _p = \Delta _c $. For this case, we find that the maximal
absorption peak is always located below the the probe absorption of
the standard three-level EIT.

\section{The probe absorption under steady-state condition}
Then we examined the probe absorption under the steady-state
condition, i.e., $\partial\rho_{ij}/\partial t=0$. The general
steady-state solutions for Eqs. (\ref{eq2})-(\ref{eq10}) can be
derived analytically (analyzing with two resonant coupling and probe
fields), but the tedious and intricate expressions for the solutions
corresponding to the situation of an off-resonant coupling field
offers no clear physical sight. So we performance the numerical
calculation here. The present process is most clearly understood
according to the dressed states produced by the strong coupling
field $\omega_c$ as shown in Fig. \ref{fig1}(b), i.e., the
transitions $\left| 2 \right\rangle \leftrightarrow \left| 3
\right\rangle $ and $\left| 2 \right\rangle \leftrightarrow \left| 4
\right\rangle $ together with the coupling field are treated as a
total system forming the dressed states. Under the action of the
strong coherent coupling field, levels $\left| 3 \right\rangle $ and
$\left| 4 \right\rangle $ should be split into three dressed-state
levels $\left| \pm \right\rangle $ and $\left| 0 \right\rangle $.
The energy eigenstates of the dressed states for the $\Delta _c = 0$
case are written as follows

\begin{equation}
\label{eq12} \left| + \right\rangle = - \frac{\Omega _c }{\sqrt
{\Delta ^2 + 2\Omega _c^2 } }\left| 2 \right\rangle -
\frac{1}{2}\frac{\Delta - \sqrt {\Delta ^2 + 2\Omega _c^2 }}{\sqrt
{\Delta ^2 + 2\Omega _c^2 } }\left| 3 \right\rangle  +
\frac{1}{2}\frac{\Delta + \sqrt {\Delta ^2 + 2\Omega _c^2 }}{\sqrt
{\Delta ^2 + 2\Omega _c^2 } }\left| 4 \right\rangle,
\end{equation}

\begin{equation}
\label{eq13} \left| 0 \right\rangle = \frac{\Delta }{\sqrt {\Delta
^2 + 2\Omega _c^2 } }\left| 2 \right\rangle - \frac{\Omega _c
}{\sqrt {\Delta ^2 + 2\Omega _c^2 } }\left| 3 \right\rangle
 + \frac{\Omega _c }{\sqrt {\Delta ^2 + 2\Omega _c^2 }
}\left| 4 \right\rangle,
\end{equation}

\begin{equation}
\label{eq14} \left| - \right\rangle = \frac{\Omega _c }{\sqrt
{\Delta ^2 + 2\Omega _c^2 } }\left| 2 \right\rangle +
\frac{1}{2}\frac{\Delta + \sqrt {\Delta ^2 + 2\Omega _c^2 }}{\sqrt
{\Delta ^2 + 2\Omega _c^2 } } \left| 3 \right\rangle  -
\frac{1}{2}\frac{\Delta - \sqrt {\Delta ^2 + 2\Omega _c^2 }}{\sqrt
{\Delta ^2 + 2\Omega _c^2 } }\left| 4 \right\rangle.
\end{equation}

The corresponding energy eigenvalues are given by $\lambda _0 = 0$
and $\lambda _\pm = \pm \sqrt {\Delta ^2 + 2\Omega _c^2 } $. And
the dressed-state transition dipole moments can be derived as:
$\mu _{ + 1} = \left\langle + \right|P\left| 1 \right\rangle = \mu
_p $, $\mu _{01} = \left\langle 0 \right|P\left| 1 \right\rangle =
0$, $\mu _{ - 1} = \left\langle - \right|P\left| 1 \right\rangle =
\mu _p $, where $P = \mu _p (\left| 3 \right\rangle \left\langle 1
\right| + \left| 4 \right\rangle \left\langle 1 \right|)$
represents the polarization operator of the probe transition.

Based on the above analysis, it is straightforwardly shown that the
transition between dressed-levels $\left| 0 \right\rangle $ and
$\left| 1 \right\rangle $ is forbidden due to the couplings of
$\left| 1 \right\rangle \leftrightarrow \left| 3 \right\rangle $ and
$\left| 1 \right\rangle \leftrightarrow \left| 4 \right\rangle $. As
shown in Fig.\ref{fig7}, we can observe two symmetric absorption
peaks with same amplitude on the probe spectra in the case of
$\Delta _c = 0$ because of the relation $\mu _{ + 1} = \mu _{ - 1}
$. The two absorption peaks correspond respectively to the
dressed-state transitions $\left| 1 \right\rangle \leftrightarrow
\left| + \right\rangle $ and $\left| 1 \right\rangle \leftrightarrow
\left| - \right\rangle $, so the peaks are always located at $\Delta
_p = \pm \sqrt {\Delta ^2 + 2\Omega _c^2 } $, and the difference
between them becomes larger with increasing $\Omega _c $. Since the
transition $\left| 1 \right\rangle \leftrightarrow \left| 0
\right\rangle $ in the dressed-state picture is
interference-cancelled out and this interference always makes a
negative contribution to the probe absorption, thus it is easy to
understand that the probe absorption at the line center can be
approximately equal to zero, i.e., the probe absorption can be
completely suppressed, so that the absorption peak at the line
center is too small to be observed. As shown in Fig.\ref{fig8}, we
plot the probe absorption profile versus the probe detuning $\Delta
_p $. One can find that the probe absorption spectra are always
approximately equal to zero within the range of the splitting
$2\Delta$ between two tunnel-coupled levels ($\left| 3 \right\rangle
$ and $\left| 4 \right\rangle )$. This means that if the two-photon
Raman resonance is satisfied ($\Delta_p=\Delta_c$), we can observe
the transparency window without requiring one- or two-photon
detunings exactly vanish.

\begin{figure}\label{fig7}
\includegraphics[width=7cm]{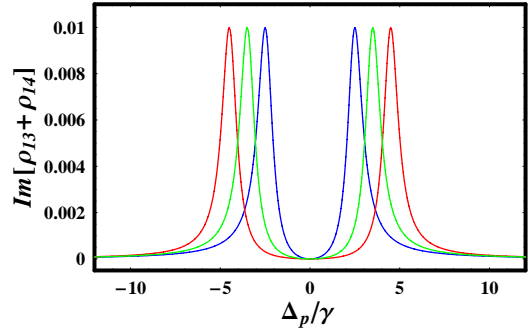}
\caption{\label{fig7} Imaginary part of the susceptibility of the
probe transition as a function of dimensionless detuning
$\Delta_{p}/\gamma$ under the condition of steady-state process with
$\Omega_{c}=2\gamma$ (blue curve), $\Omega_{c}=3\gamma$ (green
curve) and $\Omega_{c}=4\gamma$ (red curve). Other parameters used
are $\Omega_p=0.01\gamma$, $\Delta=\gamma$, $\Delta_c=0$,
$\gamma_{3l}=\gamma_{4l}=0.4\gamma$,
$\gamma_{14}^{dph}=\gamma_{24}^{dph}=\gamma_{13}^{dph}=\gamma_{23}^{dph}=0.1\gamma$.}
\end{figure}

\begin{figure}\label{fig8}
\includegraphics[width=8cm]{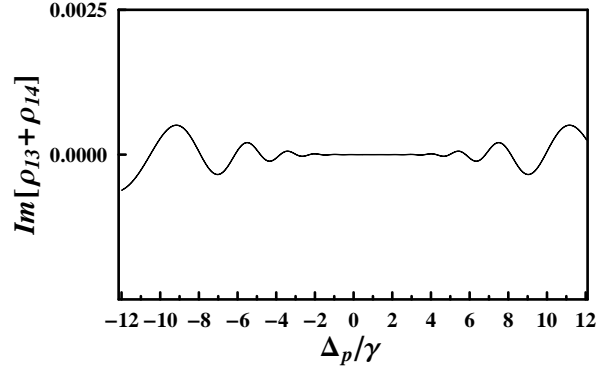}
\caption{\label{fig8} Imaginary part of the susceptibility of the
probe transition as a function of dimensionless detuning
$\Delta_{p}/\gamma$ under the condition of steady-state process for
the Raman resonance ($\Delta_c=\Delta_p$) with the parameters being
$\Omega_p=0.01\gamma$, $\Omega_c=4\gamma$, $\Delta=\gamma$,
$\gamma_{3l}=\gamma_{4l}=0.4\gamma$,
$\gamma_{14}^{dph}=\gamma_{24}^{dph}=\gamma_{13}^{dph}=\gamma_{23}^{dph}=0.1\gamma$.}
\end{figure}

\section{Discussions and Conclusion}
We now give a brief discussion on the practical setup. The
asymmetric semiconductor coupled double QW structure, considered
here, consists of 10 pairs of a 51-monolayer (145{\AA}) thick wide
well and a 35-monolayer (100{\AA})thick narrow well, separated by a
Al$_{0.2}$Ga$_{0.8}$ As buffer layer, as shown in Fig. \ref{fig1}.
The probe field propagates in the polarization direction. As in the
report in Ref.\cite{33}, we consider a transverse magnetic polarized
probe incident at an angle of 45 degrees with respect to the growth
axis so that all transition dipole moments include a factor
$1/\sqrt{2}$ as subband transitions are polarized along the growth
axis. In the sections III and IV, all the calculations have assumed
that the parameters $\Omega _c $, $\Omega _p $, $\Delta _c $,
$\Delta _p $, $\Delta$, $\gamma _{ij} $, $\gamma _{ij}^{dph} $ are
defined as units of $\gamma $ in this present paper. By setting
$\gamma=1.25$THz, the corresponding parameters value in the present
paper are respectively as $\gamma_{3l}=\gamma_{4l}=0.5$THz,
$\gamma_{14}^{dph}=\gamma_{24}^{dph}=\gamma_{13}^{dph}=\gamma_{23}^{dph}=0.125$THz,
$\Omega_p=0.0125$THz, The maximal value of $\Omega_c$ is about 5THz,
which can be realized with current experimental technology
\cite{33}.

In conclusion, in the present paper, we have analyzed and discussed
the probe absorption properties in a tunnel-coupled double quantum
wells under the conditions for the transient regime and the
steady-state process by using the detailed numerical simulations
based on the density matrix equations (\ref{eq2})-(\ref{eq10}).
Besides, we analytically analyze probe absorption mechanism for the
steady-state process in the dressed-state picture. In the
steady-state process, we show that the probe absorption can be
largely suppressed with the coupling field resonance ($\Delta_c=0$),
and be eliminated completely under the condition of Raman resonance
($\Delta_p=\Delta_c$) within the range of the $2\Delta$. In the
transient-state process, we show that the magnitude of the probe
absorption at the line center of the probe transition $\Delta_p=0$
can be greatly suppressed. Comparing with the EIT scheme in the
three-Level QW structures\cite{12}-\cite{14}, our results show that
the transparency hole is much broader due to the fact that the
existence of the tunnel-coupling splitting $2\Delta$. Besides, we do
not require that the one- or two-photon detunings is exactly zero.
As a result, the present investigation confirm the possibility of
obtaining EIT in solids by using asymmetric tunnel-coupled QW. Our
calculations also provide a guideline for the optimal design to
achieve very fast and low-threshold all-optical switches in such
semiconductor systems which is much more practical than that in
atomic system because of its flexible design and the controllable
coherent coupling strength.

The research is supported in part by National Fundamental Research
Program of China 2005CB724508, by National Natural Science
Foundation of China under Grant Nos. 10704017, 10634060, 90503010
and 10575040.

\end{document}